\documentclass[11pt, a4paper]{article}
\usepackage{tikz}
\usetikzlibrary{snakes}
\usepackage[utf8]{inputenc} 
\usepackage{textcomp} 
\usepackage{flafter}  

\usepackage{amsmath,amssymb}  
\usepackage{bm}  
\usepackage{authblk}
\usepackage{memhfixc}  
\usepackage{listings}
\numberwithin{equation}{section}
\begin{document}
\title{Exact Green's Function of the reversible diffusion-influenced reaction for an isolated pair in 2D}
\author{Thorsten Pr\"ustel} 
\author{Martin Meier-Schellersheim} 
\affil{Laboratory of Systems Biology\\National Institute of Allergy and Infectious Diseases\\National Institutes of Health}
\maketitle
\let\oldthefootnote\thefootnote 
\renewcommand{\thefootnote}{\fnsymbol{footnote}} 
\footnotetext[1]{Email: prustelt@niaid.nih.gov, mms@niaid.nih.gov} 
\let\thefootnote\oldthefootnote 
\abstract
{
We derive an exact Green's function of the diffusion equation for a pair of spherical interacting particles in 2D subject to a back-reaction boundary condition.
}

\section{Introduction}
Green's functions (GF) for an isolated pair of molecules play an important role in the theory of diffusion-influenced reactions for several reasons.
First, the solution for any other initial distribution can be calculated when this GF is known \cite{Agmon:1984, Agmon:1990p10, kimShin:1999}.
Second, the GF can be used to derive important other quantitities, for instance the survival probability and the time-dependent rate coefficient \cite{Agmon:1984, Agmon:1990p10, kimShin:1999}.
Third, the GF may be used in particle-based stochastic simulation algorithms to enhance the efficiency of naive Brownian dynamics simulations, cp for instance \cite{vanZon:2005p340}.
Finally, the knowledge of an exact analytic expression permits to validate newly devised stochastic simulation algorithms \cite{vanZon:2005p340}.

Exact analytic expressions have been derived for the GF of an isolated pair that can undergo a reversible reaction in 1D and 3D \cite{Agmon:1984, Agmon:1990p10, kimShin:1999}.
To the best of our knowledge, no exact expression has been presented so far for the 2D case. However, in particular with regard to cell biological applications, 
an accurate theoretical treatment of two-dimensional diffusion-influenced reactions provides the basis for a better understanding of important processes like receptor 
clustering on cell membranes \cite{alberts2008molecular}. 
 
To derive the GF in 2D we consider an isolated pair of two spherical particles $A$ and $B$ with diffusion constants $D_{A}$ and $D_{B}$, respectively.
The particles may associate when their separation equals the "encounter distance" $a$ to form a bound molecule $AB$. When bound, the molecules may dissociate again to form an unbound pair $A+B$.
Such a system may be described as the diffusion of a point-like particle with diffusion constant $D=D_{A}+D_{B}$ around a static sphere with radius $a$. In this picture, reactions are introduced by imposing boundary conditions at the sphere's surface. The association reaction is described by the radiation boundary condition that is characterized by an intrinsic association constant $\kappa_{a}$.
The radiation boundary condition is used to describe irreversible association. To take into account reversible reactions, i.e. dissociations also, the radiation boundary condition has to be generalized to the back-reaction boundary condition that involves an additional intrinsic dissociation constant $\kappa_{d}$. 

We consider the probability density function $g(r, t\vert r_{0})$ for the probability to find the particles at a distance equal to $r$ at time $t$ given that the distance was inititially $r_{0}$ at time $t=0$. The time evolution of $g(r, t\vert r_{0})$ is governed by the 2D diffusion equation   
\begin{equation}\label{2DEinsteinDiffEq}
\frac{\partial}{\partial t} g(r, t \vert r_{0})= D(\frac{\partial^{2}}{\partial r^{2}}+\frac{1}{r}\frac{\partial}{\partial r})g(r , t \vert r_{0}), \quad r \geq a
\end{equation}
The diffusion equation has to be completed by specifying boundary conditions. Together with the following initial
\begin{equation}\label{initial_bc}
2\pi rg(r, t_{0}\vert r_{0}, t_{0})=\delta(r-r_{0})
\end{equation}
and boundary condition
\begin{equation}\label{inf_bc}
g(r \rightarrow\infty, t\vert r_{0}, t_{0})=0
\end{equation}
equation \eqref{2DEinsteinDiffEq} (if $a = 0$) is equivalent to the free-space diffusion equation in 2D with the familiar solution
\begin{equation}\label{freeGF2D}
g_{\text{free}}(r, t \vert r_{0}) = \frac{1}{4\pi D t}
 e^{-(r^{2}+r^{2}_{0})/4Dt}I_{0}(\frac{rr_{0}}{2Dt}) 
\end{equation}
also known as the free-space Green's function. 

However, as described previously, the PDF we are interested in is only defined for $r \geq a > 0$ and one has to impose a boundary condition for $r=a$ specifying the behavior at the encounter distance. 
We will discuss the following case \cite{Agmon:1984, kimShin:1999}:
\begin{equation}\label{BRBC}
2\pi a D \frac{\partial}{\partial r}g(r, t\vert r_{0})\vert_{r=a} =\kappa_{a}g(r=a, t\vert r_{0}) -\kappa_{d}[1-S(t\vert r_{0})].
\end{equation} 
Here $S(t\vert r_{0})$ denotes the survival probability that a pair of molecules with initial distance $r_{0}$ survives by time $t$
\begin{equation}
S(t\vert r_{0}) = 1 - \int^{t}_{0}2 \pi a D \frac{\partial}{\partial r}g(r, t^{\prime}\vert r_{0})\vert_{r=a} dt^{\prime}
\end{equation}
Following \cite{carslaw1986conduction}, we make the following ansatz for the Laplace transform of the Green's function that satisfy the backreaction (BR) boundary condition 
\begin{equation}\label{laplaceAnsatz}
\tilde{g}(r, q \vert r_{0}) = \tilde{g}_{\text{free}}(r, q \vert r_{0}) +\tilde{g}_{\text{BR}}(r, q \vert r_{0}). 
\end{equation}
Here 
\begin{equation}\label{laplaceFree}
\tilde{g}_{\text{free}}(r, q \vert r_{0}) = \frac{1}{2\pi D}\left\{\begin{array}{lr}
 I_{0}(qr_{0}) K_{0}(qr)&\mbox{$ r  >  r_{0} $} \\
 I_{0}(qr) K_{0}(qr_{0}) &\mbox{$r  <  r_{0}$}  
\end{array}\right .
\end{equation}
is the Laplace transform of the free-space Green's function \eqref{freeGF2D}. The variable $q$ is defined by 
\begin{equation}
 q:=\sqrt{\tfrac{p}{D}} ,
\end{equation}
 where $p$ denotes the Laplace domain variable. The part $\tilde{g}_{\text{BR}}$ that takes into account the boundary condition is a solution of the Laplace transformed 2D diffusion equation 
\begin{equation}\label{laplaceDE}
\frac{d^{2}\tilde{g}_{\text{BR}}}{dr^{2}} + \frac{1}{r}\frac{d\tilde{g}_{\text{BR}}}{dr} - q^{2} \tilde{g}_{\text{BR}} = 0.
\end{equation}

The general solution to \eqref{laplaceDE} is $AK_{0}(qr) + BI_{0}(qr)$, where $I_{0}(x), K_{0}(x)$ refer to the modified Bessel functions of first and second kind, respectively, and of order zero \cite{abramowitz1964handbook}.
Because we require $\lim_{x\rightarrow \infty}\tilde{g}_{\text{BR}}\rightarrow 0 $, and $\lim_{x\rightarrow\infty}I_{0}(x)\rightarrow \infty$, the coefficient $B$ has to vanish and hence, 
\begin{equation}\label{laplaceBR}
\tilde{g}_{\text{BR}}(r, q\vert r_{0}) = A(q, r_{0}) K_{0}(qr).
\end{equation}
$A(q, r_{0})$ is determined by the requirement that the complete Green's function \eqref{laplaceAnsatz} satisfies the Laplace transformed backreaction boundary condition, cp. \eqref{BRBC} 
\begin{equation}\label{laplaceBC}
\frac{\partial}{\partial r}\tilde{g}_{\text{BR}}(r, q\vert r_{0})\vert_{r=a} = h \tilde{g}_{\text{BR}}(r, q\vert r_{0})\vert_{r=a} - \kappa_{d} p^{-1} \frac{\partial}{\partial r}\tilde{g}_{\text{BR}}(r, q\vert r_{0})\vert_{r=a}, 
\end{equation}
where we have defined $h:=\frac{\kappa_{a}}{2\pi a D}$.
Using \eqref{laplaceAnsatz}, \eqref{laplaceFree}, \eqref{laplaceBR}, \eqref{laplaceBC} and 
\begin{eqnarray}
I^{\prime}_{0}(x) &=& I_{1}(x), \\
K^{\prime}_{0}(x) &=& -K_{1}(x)
\end{eqnarray}
and defining $\kappa_{D}:=\frac{\kappa_{d}}{D}$, we obtain
\begin{equation}\label{laplaceSolution}
\tilde{g}_{\text{BR}}(r, q\vert r_{0})  = \frac{1}{2\pi D} \frac{(q^{2}+\kappa_{D})I_{1}(qa) - hqI_{0}(qa)}{(q^{2}+\kappa_{D})K_{1}(qa) + hqK_{0}(qa)}K_{0}(qr_{0})K_{0}(qr).
\end{equation}

The inversion theorem for the Laplace transformation can be applied to find the corresponding expression in the time domain
\begin{equation}\label{inversionFormula}
g_{\text{BR}}(r, t \vert r_{0}) = \frac{1}{2\pi i} \int^{\gamma+i\infty}_{\gamma-i\infty} e^{pt}\,\tilde{g}_{\text{BR}}(r, q\vert r_{0} )dp.
\end{equation}
To calculate the Bromwich contour integral \eqref{inversionFormula} we first note that $\tilde{g}_{\text{BR}}$ has a branch point at $p=0$. Therefore, we use the contour of Figure 1 with a branch cut along the negative real axis, cp. \cite{carslaw1986conduction}.  Furthermore, because the integrand has no poles within and on the contour  \cite{erdelyiKermack:1945} and because the contribution from the small circle around the origin vanishes, we obtain 
\begin{eqnarray}
0 &=& \oint e^{pt}\,\tilde{g}_{\text{BR}}(r, q\vert r_{0} )dp 
= \int^{\gamma+i\infty}_{\gamma-i\infty} e^{pt}\,\tilde{g}_{\text{BR}}(r, q\vert r_{0} )dp + \nonumber\\
&+& \int_{\mathcal{C}_{2}} e^{pt}\,\tilde{g}_{\text{BR}}(r, q\vert r_{0} )dp + \int_{\mathcal{C}_{4}} e^{pt}\,\tilde{g}_{\text{BR}}(r, q\vert r_{0} )dp.
\end{eqnarray}
Thus, it remains to calculate the integrals $\int_{\mathcal{C}_{2}}, \int_{\mathcal{C}_{4}}$.

We now choose
\begin{equation}
p = D x^{2} e^{\pi i}
\end{equation}
and use \cite{carslaw1986conduction}
\begin{eqnarray}
I_{n}(xe^{\pm \pi i/2}) &=& e^{\pm n\pi i/2} J_{n}(x), \\
K_{n}(xe^{\pm \pi i/2}) &=& \pm\frac{1}{2}\pi i e^{\mp n\pi i/2} [-J_{n}(x) \pm i Y_{n}(x)].
\end{eqnarray}
$J_{n}(x), Y_{n}(x)$ denote the Bessel functions of first and second kind, respectively \cite{abramowitz1964handbook}.
It follows that
\begin{equation}
\int_{\mathcal{C}_{2}}e^{pt}\,\tilde{g}_{\text{BR}}(r, q\vert r_{0} )dp=\frac{i}{2}\int^{\infty}_{0}e^{-Dx^{2}t}H^{(2)}_{0}(xr)H^{(2)}_{0}(xr_{0})F(x) x dx.
\end{equation}
Here $H^{(2)}_{n}(x) := J_{n}(x) - i Y_{n}(x) $ denotes the Bessel function of third kind (also referred to as Hankel function) \cite{abramowitz1964handbook}
and we have defined 
\begin{equation}
F(x):=\frac{\alpha(x)[\alpha(x)+i\beta(x)]}{\alpha(x)^{2} + \beta(x)^{2}}
\end{equation}
and
\begin{eqnarray}
\alpha(x) &:=& ( -x^{2} + \kappa_{D})J_{1}(xa) - hxJ_{0}(xa) \\
\beta(x) &:=& ( -x^{2} + \kappa_{D})Y_{1}(xa) - hxY_{0}(xa) 
\end{eqnarray}

To evaluate the integral along the contour $\mathcal{C}_{4}$ we choose $p = Dx^{2}e^{-i\pi}$ and after an analogous calculation one finds that
\begin{equation}
\int_{\mathcal{C}_{2}}e^{pt}\,\tilde{g}_{\text{BR}}(r, q\vert r_{0} )dp=-\overline{\int_{\mathcal{C}_{4}}e^{pt}\,\tilde{g}_{\text{BR}}(r, q\vert r_{0} )dp}
\end{equation}
where $\overline{\int_{\mathcal{C}_{4}}}$ means complex conjugation.
Thus, one arrives at
\begin{eqnarray}
g_{\text{BR}}(r, t \vert r_{0}) &=& \frac{1}{2\pi i} \int^{\gamma+i\infty}_{\gamma-i\infty} e^{pt}\,\tilde{g}_{\text{BR}}(r, q\vert r_{0} )dp \\
&=& -\frac{1}{\pi} \Im\left(\int_{\mathcal{C}_{2}}e^{pt}\,\tilde{g}_{\text{BR}}(r, q\vert r_{0} )dp\right) \\
&=& -\frac{1}{2\pi}\int^{\infty}_{0}e^{-Dx^{2}t} \frac{\alpha(x)[\alpha(x)\Omega(x) + \beta(x)\Pi(x)]}{\alpha(x)^{2} + \beta(x)^{2}}   xdx
\end{eqnarray}
where we have defined 
\begin{eqnarray}
\Omega(x) := J_{0}(xr)J_{0}(xr_{0}) - Y_{0}(xr)Y_{0}(xr_{0}) \\
\Pi(x) : = Y_{0}(xr)J_{0}(xr_{0}) + J_{0}(xr)Y_{0}(xr_{0}).
\end{eqnarray}

Next, we use the fact that the free-space Green's function may be written as
\begin{equation}
g_{\text{free}}(r, t\vert r_{0}) = \frac{1}{2\pi}\int^{\infty}_{0}e^{-Dx^{2}t}J_{0}(xr)J_{0}(xr_{0})x dx
\end{equation}
to arrive at the exact Green's function in the time domain
\begin{equation}
g(r, t\vert r_{0} )=\frac{1}{2\pi}\int^{\infty}_{0}e^{-Dx^{2}t}T_{0}(xr)T_{0}(xr_{0})xdx
\end{equation}
with
\begin{equation}
T_{0}(xr):=\tfrac{ J_{0}(rx)[(x^{2}-\kappa_{D})Y_{1}(xa) + hx Y_{0}(xa)] - Y_{0}(rx)[(x^{2}-\kappa_{D})J_{1}(xa) + h xJ_{0}(xa)]}{\lbrace[(x^{2}-\kappa_{D})J_{1}(xa) + h x J_{0}(xa)]^{2} + [(x^{2}-\kappa_{D})Y_{1}(xa) + hxY_{0}(xa)]^{2}\rbrace^{1/2}}. 
\end{equation}
Note that the limit $\kappa_{d}\rightarrow 0$ one recovers the known Green's function \cite{carslaw1986conduction} for the irreversible case with radiation boundary condition.

\begin{tikzpicture}[scale=2]
\node[at={(1.2,2)}]{Figure 1};
\draw (-2.4, -0.08) -- (2.4, -0.08)node[anchor=north]{$\Re$};
\draw (-0.4, -2.4) -- (-0.4, 2.4) node[anchor=west]{$\Im$};
\draw [ snake=snake](-2.4, -0.08) -- (-0.35, -0.08);
\draw [thick, ->](0., -2.16) -- (0., 2.) node[near end, anchor=west]{$\mathcal{C}_{1}$};
\draw [<-,thick](-2, 0) arc (180:90:2);
\draw [->, thick](-2., 0.) -- (-0.55, 0.) node[midway, anchor=south]{$\mathcal{C}_{2}$};
\draw [<-, thick](-0.58, -0.15)  arc  (-160:160:0.2);

\draw [->, thick](-2, -0.15) arc (180:270:2);
\draw[<-, thick] (-2., -0.15) -- (-0.55, -0.15) node[midway, anchor=north]{$\mathcal{C}_{4}$};
\end{tikzpicture}

\subsection*{Acknowledgments}
This research was supported by the Intramural Research Program of the NIH, National Institute of Allergy and Infectious Diseases. 

We would like to thank Bastian R. Angermann and Frederick Klauschen for helpful and stimulating discussions.

\bibliographystyle{plain} 
\bibliography{GreenBD}

\end{document}